\documentclass[prl,showpacs,twocolumn,superscriptaddress,floatfix]{revtex4}
\usepackage{amsmath}
\usepackage{graphicx}
\usepackage{float}
\usepackage{amssymb}
\usepackage{times}
\DeclareGraphicsExtensions{.eps,.ps}

\begin{document}
\title{Impurity-induced frustration in correlated oxides}

\author{Cheng-Wei Liu}
\affiliation{Department of Physics, National Taiwan University,
  Taipei, Taiwan 106} 

\author{Shiu Liu}
\affiliation{Department of Physics and Astronomy, University of
  California, Irvine, California 92697}

\author{Ying-Jer Kao}
\email{yjkao@phys.ntu.edu.tw}
\affiliation{Department of Physics, National Taiwan University,
  Taipei, Taiwan 106} 

\author{A. L. Chernyshev }
\email{sasha@uci.edu}
\affiliation{Department of Physics and Astronomy, University of
  California, Irvine, California 92697}

\author{Anders W. Sandvik}
\affiliation{Department of Physics, Boston University, 590
  Commonwealth Ave., Boston, Massachusetts 02215}
\affiliation{Department of Physics, National Taiwan University,
  Taipei, Taiwan 106} 

\date{\today}

\begin{abstract}
Using the example of Zn-doped La$_2$CuO$_4$, we 
demonstrate that a spinless impurity doped into a non-frustrated 
antiferromagnet can induce substantial frustrating interactions
among the spins surrounding it. This counterintuitive result is 
the key to resolving discrepancies between experimental data and earlier 
theories.  Analytic and quantum Monte Carlo studies of the impurity-induced 
frustration are in a close accord with each other and experiments.
The mechanism proposed here should be common to other correlated 
oxides as well.
\end{abstract}
\maketitle

Impurities are known to be an effective tool to locally perturb
quantum systems,  
thereby revealing important information about their microscopic
interactions and correlations \cite{Haas}.  
A well studied example of a strongly correlated quantum system in
which effects of such impurity doping can be  
investigated is La$_2$CuO$_4$---one of the most important cuprate
superconductor parent compounds. In its pristine  
form, this material is a two-dimensional (2D) spin-$\frac12$
Heisenberg antiferromagnet (AF) \cite{CHN}. It is  
believed that the substitution of Cu$^{2+}$ ($S=\frac12$) ions by
spinless Zn$^{2+}$ represents a good 
realization of the site-diluted Heisenberg hamiltonian
\cite{Greven,Carretta,Sandvik1,us_YC1}. In this Letter, we  
demonstrate that there exists a significant {\it qualitative}
correction to the dilution picture. Impurities can  
induce substantial \textit{frustrating interactions} between nearby
spins. Not only does this effect explain  
discrepancies between experimental data and the dilution-only theories
for La$_2$Cu$_{1-x}$Zn$_x$O$_4$,  
but it may also be important for a variety of other phenomena in
diluted magnets and doped Mott insulators.  
Our mechanism for such an effect should be common to many
charge-transfer insulators, including oxides  
of transition metals. 

We propose that the presence of extra degrees of freedom due to oxygen
orbitals necessarily results in frustrating terms in  
the corresponding low-energy spin hamiltonian of the 
Zn-doped system, which are absent in the dilution-only
models. Utilizing quantum Monte Carlo (QMC) and  
analytic $T$-matrix approaches, we calculate the doping dependence of
the staggered magnetization for such  
a low-energy model.
We show that  
this model, with the parameters appropriate for the CuO$_2$ planes
given by a three-band Hubbard model calculation,  
naturally explains experimental data. 
 
\begin{figure}[b]
\includegraphics[angle=0,width=8cm,clip=true]{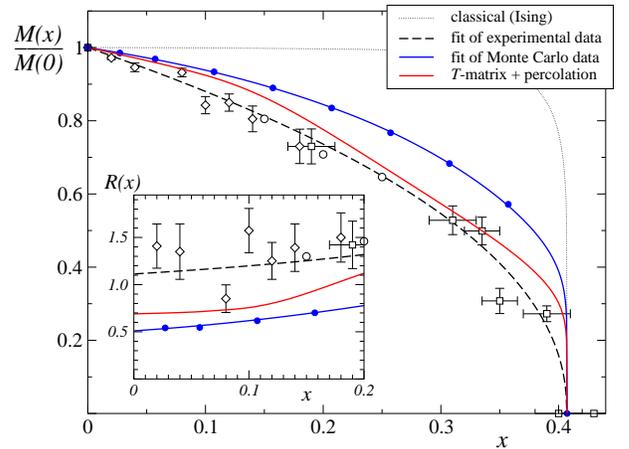}
\caption{(Color online). The on-site magnetic moment $M(x)$ 
per Cu normalized by its undoped value vs Zn doping $x$. Symbols show neutron 
scattering \cite{Greven} (squares) and NQR \cite{Carretta} (diamonds
and circles)  
data, with dashed lines their best fits. The dotted line is the classical 
($S\!\rightarrow\!\infty$) result. Solid lines are the QMC \cite {Sandvik1} and 
the $T$-matrix results \cite{us_YC1}, respectively. Inset: slope function 
(see text) vs $x$ \cite{footnote}}.
\label{dil2}
\vskip-3mm
\end{figure}

{\it Experiments and theories.}---Comprehensive
studies of the problem of La$_2$CuO$_4$  
diluted by spinless Zn impurities have been performed using neutron
scattering, magnetometry, and NMR (NQR)  
on the experimental side \cite{Greven,Carretta}, and QMC and
$T$-matrix approaches of the diluted  
Heisenberg model on the theoretical side
\cite{Sandvik1,Kato,us_YC1}. These studies allow 
for extensive cross-checks. The unbiased QMC data agree with the
$T$-matrix results closely  
up to $x\simeq 15\%$, supporting the validity of the latter in the
low-doping regime \cite{us_YC1,Sandvik1}. However, there are serious
discrepancies between theoretical and experimental
results. Fig.~\ref{dil2} shows the average magnetic moment $M$ per Cu
site versus  
the Zn doping fraction $x$. The experimental data are always below the
theoretical  
curves. The slope 
\begin{equation}
\label{slope1}
R(x)=\frac{1}{x}\left(1-\frac{M(x)}{M(0)}\right), 
\end{equation} 
at small $x$ represents the rate at which the order parameter $M$ is
suppressed by individual  
impurities due to enhanced quantum fluctuations. The inset of
Fig.~\ref{dil2} shows a large  
discrepancy---a factor of approximately two---between the theoretical
and experimental results.  
This indicates that the dilution-only theory significantly
underestimates the impact of the  
impurity on the quantum spin background. 

\begin{figure}[t]
\includegraphics[angle=0,width=8cm]{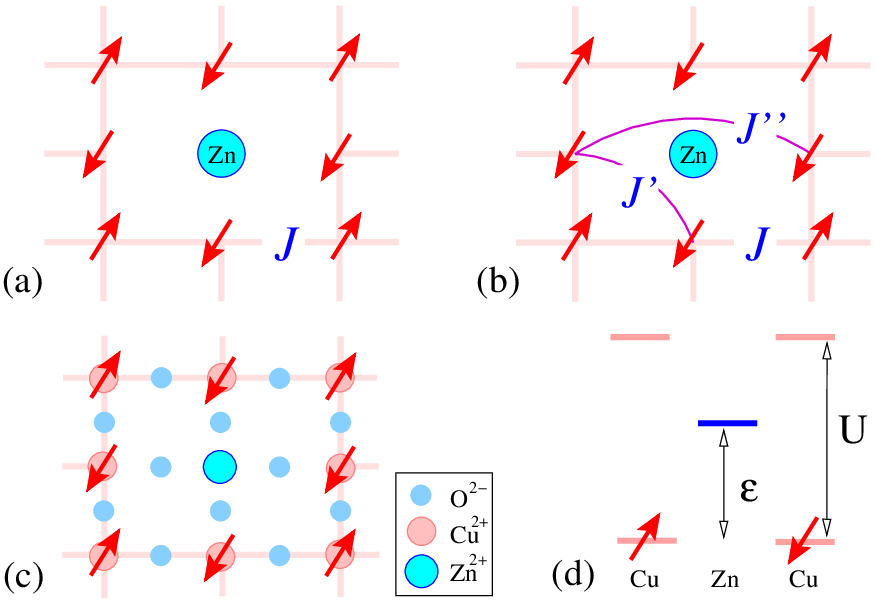}
\caption{(Color online) AF bonds in the (a) dilution-only model,
(b) model with impurity-induced frustration (two out of six frustrating 
bonds are shown). (c) The CuO$_2$ plane with a Zn impurity. (d) Schematic
local level diagram in the impurity-doped system.} 
\label{spinsZn}
\vskip-3mm
\end{figure}

One might attempt to explain the disagreement by suggesting that
longer-range ($J_2$, $J_3$, etc.), and ring-exchange  
interactions should be included in the model for the undoped CuO$_2$
plane. Such terms are generally present in the low-energy  
spin models derived from the Hubbard model \cite{Hubbard}, and they 
do lead to a reduction of $M$. However, since 
the order is suppressed already in the undoped system, 
this mechanism is unlikely to enhance fluctuations 
specifically due to dilution. Using an expansion of $M$ in the 
dilution fraction $x$ and in the extended interaction $J_2$, 
one obtains at small $x$ and $J_2$:
\begin{equation}
R(J_2)\approx
R(0)\left(1+A\frac{J_2}{J}\right),
\end{equation}
where $A\alt 1$ \cite{Poilblanc} 
and $R(0)$ is the theory slope from Fig.~\ref{dil2}. Thus, a
large correction to the slope  
of the $M(x)/M(0)$ curve in the extended model is only possible if
$J_2\!\sim\!J$, which is beyond the realistic range in the cuprates 
where $J_2/J$ is at most of the order of 10\% \cite{Delannoy}. 
A recent study \cite{Delannoy} has shown that while extended 
interactions are important for explaining the lower {\it absolute} 
value  of the staggered magnetization, they are not able to explain 
the large initial slope in the $M(x)/M(0)$ dependence. Thus,
one must seek another explanation.

{\it Extra interactions.}---The dilution picture seems natural
for modeling the replacement of a magnetic Cu site by a magnetically 
inert Zn; see Fig.~\ref{spinsZn}(a). 
However, for the dilution-only picture to be valid,   
the Zn-site must remain {\it electronically} inert at energy 
scales up to the Hubbard $U$. Since it is 
the three-band Hubbard model that describes the real 
CuO$_2$ plane and other transition-metal oxides \cite{Emery},
the states on the oxygen orbitals also become important.
They hybridize to Zn and remain involved in 
virtual hoppings between surrounding Cu-sites, see Fig.~\ref{spinsZn}(c),
 facilitating extra superexchange couplings that connect 
further neighbor Cu-sites. 
Thus, the spinless impurity, in effect,  
leads to a cage of frustrating interactions around itself, with four
$J^\prime_{Zn}$ and two $J^{\prime\prime}_{Zn}$; see Fig.~\ref{spinsZn}(b). 
Qualitatively, the impurity-doped system
is {\it not} equivalent to the site-diluted Hubbard  
model with electronically inert impurity sites, but
rather to the $t$-$\varepsilon$-$U$ model,  
\begin{equation} 
\label{Heff}
H=-t\sum_{ij,\sigma} c^\dag_{i,\sigma}c_{j,\sigma}+
\varepsilon\sum_{l,\sigma} 
n_{l,\sigma}^{Zn}+U\sum_i n_{i,\uparrow} n_{j,\downarrow} \ ,
\end{equation} 
a visual representation of which is given in
Fig.~\ref{spinsZn}(d). The $t$-$U$ part is the usual Hubbard model,  
which at half-filling reduces to the Heisenberg model at $t^2/U$
order. The higher-order terms  are negligible  
($\sim\! t^4/U^3$) if $t\!\ll\! U$. The distinct physics is brought into
play by the model (\ref{Heff}) when the energy cost at  
the impurity site $\varepsilon$ is less than the Hubbard gap. In that
case, virtual transitions through the  
impurity level will cause superexchange interactions of order of $\sim\!
t^4/\varepsilon^3$. Taking $\varepsilon=U/2$  
and $U/t=10$ for an estimate leads to $J^\prime_{Zn}/J\!\sim\!
(t/U)^2(U/\varepsilon)^3\!\sim\! 0.1$. The total impact of the  
impurity-induced frustrating interactions per impurity is then
$J^\prime_{tot}\!=\!4J^\prime_{Zn}\!+\!2J^{\prime\prime}_{Zn}\!\sim\! 0.6 J$,  
which, as we will show below, is enough to explain the discrepancy
between dilution-only theory and experiments. As is discussed  
above, the corresponding Hubbard terms of the 4th order are 
smaller and do not disturb the order  
parameter specifically due to dilution.

For the realistic values of the CuO$_2$ plane parameters,   
mapping of the three-band Hubbard model to the single-band one 
can be done using the cell-perturbation approach \cite{3band} 
which does not require the smallness of the Cu-O hopping 
$t_{pd}$ with respect to the
charge-transfer gap $\Delta$ \cite{Zhang_Rice}. In this approach, locally 
hybridized states on Cu and surrounding O's are diagonalized 
exactly and the three-band model becomes a 
``multi-orbital'' Hubbard model with the effective ``Cu'' states  
 connected by effective hoppings. 
Since the lowest states in the multi-orbital Hubbard model 
are the same as in the single-band one ({\it i.e.} the lowest two-hole state 
is the Zhang-Rice-like singlet) the equivalence of the two models can be 
justified \cite{3band}. In this approach, 
even if Zn is inert electronically, 
the remaining O-like states in the Zn-O$_4$ cluster 
can facilitate couplings between neighboring  
Cu spins, Fig.~\ref{spinsZn}(c).

\begin{figure}[t]
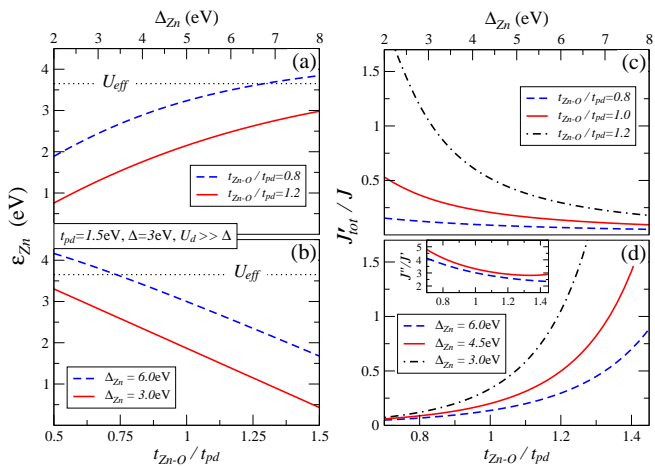

\begin{center}
\includegraphics[width=0.495\columnwidth]{new_fig3ab}  
\includegraphics[width=0.49\columnwidth]{new_fig3cd}
\caption{(Color online) (a) and (b): energy $\varepsilon_{Zn}$ 
of the lowest state on the effective ``Zn''-site (ZnO$_4$ cluster)
vs $\Delta_{Zn}$ and $t_{Zn-O}$, respectively.
(c) and (d): $J^\prime_{tot}\!=\!4J^\prime_{Zn}\!+\!2J^{\prime\prime}_{Zn}$ vs
$\Delta_{Zn}$ (for several values of $t_{Zn-O}$) and $t_{Zn-O}$ (for
several values of $\Delta_{Zn}$), respectively. Inset (d): 
$J^{\prime\prime}_{Zn}/J^\prime_{Zn}$ vs $t_{Zn-O}$. 
In (a-d), $t_{pd}\!=\!1.5$eV, $\Delta\!=\!3$eV, $U_d\!\gg\!\Delta$.} 
\label{fig3}
\end{center}
\vskip-3mm
\end{figure}

We extend this approach to the Zn-doped 
case and perform a detailed
microscopic calculations of $J^\prime_{Zn}$ and
$J^{\prime\prime}_{Zn}$ \cite{Liu_PRB}.
First, we fix the parameters 
of the three-band model so they yield the experimental value of the
Cu-Cu superexchange $J\!\simeq\!0.13$eV \cite{3band}. Since the
electronic parameters of Zn states are not known precisely 
\cite{Plakida,Hirschfeld}, we vary them substantially as shown in 
Fig.~\ref{fig3} for a representative set of 
the three-band model parameters. Our Figs.~\ref{fig3}(a),(b) show 
how the energy $\varepsilon_{Zn}$ of the lowest effective ``Zn'' state  
depends on the  energy of the bare 
Zn-level, $\Delta_{Zn}$, and the hybridization, 
$t_{Zn-O}$, respectively. The effective Hubbard energy $U_{eff}$ 
is also shown to demonstrate the validity of the qualitative
level structure in Fig.~\ref{spinsZn}(d) and to support our model 
(\ref{Heff}). Figs.~\ref{fig3}(a),(b) show that 
the electronic levels of Zn and hybridization with them are important  
in lowering $\varepsilon_{Zn}$ and enhancing $J^\prime_{Zn}$ and
$J^{\prime\prime}_{Zn}$. 

Our Figs.~\ref{fig3}(c),(d) show the $\Delta_{Zn}$ and 
$t_{Zn-O}$ dependence of the total impurity-induced frustrating
interactions per impurity  $J^\prime_{tot}/J$.
Given the uncertainty in $\Delta_{Zn}$ (from 2-3eV, \cite{Hirschfeld},
to 5eV, \cite{Plakida}), the total frustrating effect can be
estimated to be between $J^\prime_{tot}\sim (0.2-1.0) J$. Individual 
$J^\prime_{Zn}$ and $J^{\prime\prime}_{Zn}$ are in the range of 3-15\% of $J$.
Counterintuitively, the interaction across the  
impurity ($J^{\prime\prime}_{Zn}$) 
is greater than the next-nearest neighbor interaction ($J^{\prime}_{Zn}$) 
due to partial 
cancellation of the super-exchange and the ring-like exchange
involving Zn and three Cu sites on a nearest-neighbor  
plaquette, see Fig.~\ref{spinsZn}(b). 
This results in a {\it stronger} bond between copper spins
across the Zn-site, with the ratio  
$J^{\prime\prime}_{Zn}/J^{\prime}_{Zn}\simeq 2 \sim 4$ in a wide range
of the three-band model parameters, see inset in Fig.~\ref{fig3}(d). 
Altogether, the three-band model provides support to
our idea and gives an order-of-magnitude estimate of the parameters.

{\it Low-energy model.}---With this microscopic insight, one should
model the Zn-doped CuO$_2$ plane by  
the Heisenberg model with random impurities not only 
causing more fluctuations by cutting links \cite{Greven}, 
but also connecting the nearby spins in a frustrated way. 
Thus, the effective model is: 
\begin{equation}
\label{eq:Hamiltonian}
{\cal H}=J\sum_{\langle ij\rangle } \mathbf{S}_i \cdot
\mathbf{S}_j+J^\prime_{Zn}
\sum_{\langle ij\rangle'} \mathbf{S}_i \cdot
\mathbf{S}_j+J^{\prime\prime}_{Zn}
\sum_{\langle ij\rangle''} \mathbf{S}_i \cdot \mathbf{S}_j 
\end{equation}
where the first term is the dilution-only model with $S=1/2$ spins for
all the sites except the impurity sites (where $S=0$)  
and  the sums over the $J^\prime_{Zn}$ and $J^{\prime\prime}_{Zn}$
bonds are taken around the impurity sites only, as shown in  
Fig.~\ref{spinsZn}(b). 

{\it Suppression of the order parameter.}---We have investigated the
model (\ref{eq:Hamiltonian}) by means of the analytical $T$-matrix as 
well as unbiased QMC techniques. The former is based on the
diagrammatic treatment of the corresponding linear spin-wave theory
with exact  
calculation of the scattering amplitudes off the impurities and
subsequent disorder averaging. 
The details of the approach for the   
dilution-only problem are given in Ref.~\onlinecite{us_YC1} and
results for $M(x)/M(0)$ are shown in Fig. \ref{dil2}.  
The modification of this method for
the model (\ref{eq:Hamiltonian}) concerns changes in  
the $p$- and $d$-wave scatterings off the impurities, while the
$s$-wave contribution can be shown to be unaffected by the frustrating  
terms \cite{Liu_PRB}. The advantage of this method is that both the
$x$- and $J^\prime_{Zn}$, $J^{\prime\prime}_{Zn}$-dependence  
of the order parameter can be studied systematically.

QMC simulations were performed using the stochastic series expansion
method \cite{sse} to find the staggered magnetization in lattices
with $N =L \times L$ sites:
\begin{equation}
\label{Spipi}
M_L^2 = \frac{3}{N^2} 
\Big\langle \Big(  \sum_{i} (-1)^{x_i + y_i} S_i^z \Big)^2   \Big\rangle . 
\end{equation}
The frustrating interactions generally present a serious difficulty 
due to the negative sign problem \cite{PhysRevB.41.9301}, which
becomes more serious (exponentially) with increasing number of
frustrating bonds  
and inverse temperature $\beta=J/T$. Here we focus on the system with
a single impurity, where the sign problem is  
completely local (independent of the lattice size) and, in the
range of frustration of interest here, manageable down to  
sufficiently low temperatures to draw conclusions about ground states
of relatively large lattices. 
For all lattices considered below,
$M_L$ is well converged already at $\beta=16$, and in the  
following we use this value. We assume the usual size-dependence of
the staggered magnetization \cite{PRB_37_5978,PRB_56_11678}: 
\begin{equation}
M^2_L = M^2 + \frac{m_1}{L} + \frac{m_2}{L^2} + \frac{m_3}{L^3} + \ldots,
\end{equation}
As a check, we calculated $M$ for a pure AF using the same system
sizes and $\beta\!=\!16$, which gave $M\!\approx\!0.3066$ for the 
extrapolated value, only slightly below
high-precision result obtained using much larger lattices;
$M\!=\!0.30743(1)$ \cite{ARC_0807_0682}.

\begin{figure}[t]
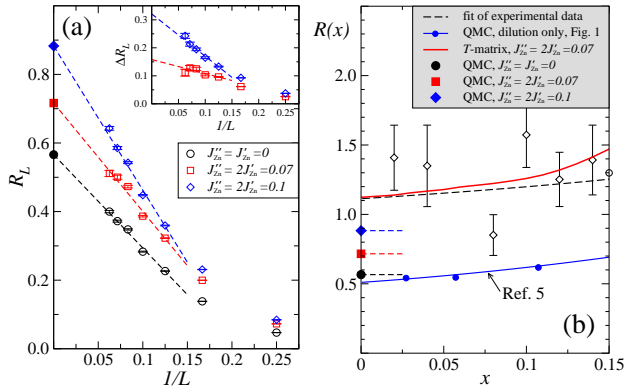

\includegraphics[width=0.45\columnwidth]{new_fig4a} \
\includegraphics[width=0.48\columnwidth]{new_fig4b}    
\caption{(Color online) (a) 
The slope $R_L$, Eq.~(\ref{RL}) vs $1/L$ in systems 
with and without frustration ($\beta\!=\!16$), solid symbols are
extrapolated values. Inset: slope increase $\Delta\! R_L$ 
due to frustration.
(b) Same as inset in Fig.~\ref{dil2}  with the $T$-matrix 
  results for $J^{\prime\prime}_{Zn}\!=\!2J^{\prime}_{Zn}\!=\!0.07J$
 and QMC extrapolations of $R(0)$ from Fig.~\ref{fig4}(a)
 for $J^{\prime\prime}_{Zn}\!=\!2J^{\prime}_{Zn}\!=0$, $0.07J$, and
 $0.1J$. QMC data provide lower bounds of the slopes. } 
\label{fig4}
\vskip-3mm
\end{figure}

We are interested in the suppression of $M(x)$ at low impurity
concentrations; $M(x)/M(0) \approx 1 - x R(0)$, where $R(0)$ is the
slope of  
the $M$ vs $x$ curve at $x\rightarrow 0$. We introduce a
finite-size analog of the slope function as 
\begin{equation}
\label{RL}
R_L = \frac{1}{x} \left[ 1 - \frac{M_{L}(1)}{M_{L}(0)} \cdot
  \frac{1}{1-x}\right], 
\end{equation}
where $L$ is the lattice size, $M_L(1)$ is the staggered magnetization
of the system doped by one impurity, and $x\!=\!1/L^2$  
for that case.  The normalization by $(1-x)$ is necessary to convert
$M$ found from (\ref{Spipi}) to weighting $M$ relative to the amount  
of magnetic sites \cite{footnote}. Fig.~\ref{fig4}(a) shows the size
dependence of the slope in both the unfrustrated $J$-only model and
the frustrated model with $J^{\prime\prime}_{Zn}\!=\! 2J^\prime_{Zn}\! =\!
0.07J$ and $0.1J$. The latter set is still below the frustration beyond 
which the sign problem becomes too serious. As expected, the frustration 
increases the slope substantially. Given the non-linearity of the data, 
there is some uncertainty in the $1/L\!\rightarrow\!0$ extrapolation. 
We here use a linear fit to $L \ge 8$ data; see Fig.~\ref{fig4}(a). Due 
to some remaining nonlinearities (which are seen clearly for sizes $L<8$), 
the extrapolation may slightly under-estimate the slope $R(0)$, and should
therefore be considered a lower bound.

To obtain a direct quantitative measure of the effect of frustration we 
consider the difference $\Delta R_L\!=\!R_L^f\!-\!R_L^0$ of the slopes 
from (\ref{RL}) for the model (\ref{eq:Hamiltonian}) with ($R^f$) and 
without ($R^0$) frustrating terms, respectively, for lattice sizes $L$ and 
several $J^\prime_{Zn}$ and $J^{\prime\prime}_{Zn}$. We perform finite-size
extrapolations to the thermodynamic limit as above for  each set of 
couplings; examples are shown in the inset of Fig.~\ref{fig4}(a).   
An unexpected finding is that two $J^{\prime\prime}_{Zn}$ bonds 
suppress the order at almost the same rate as four $J^{\prime}_{Zn}$ bonds 
of the same strength, as evidenced by both the QMC and the $T$-matrix results. 
Combining that with the systematically larger values of $J^{\prime\prime}_{Zn}$
from the three-band model calculations shows that this interaction is
particularly important.  

Fig.~\ref{fig4}(b) shows that the experimental slope $R(0)\!\approx\!1.1$ is
matched by the $T$-matrix results already at
$J^{\prime\prime}_{Zn}\!=\!2J^{\prime}_{Zn}\!=\!0.07J$ 
($J^{\prime}_{tot}\!=\!0.28J$) where we fixed the ratio to 2
according to the three-band model  
results.  QMC results from Fig.~\ref{fig4}(a), extrapolated to
$x\!\rightarrow\! 0$, are shown in  
Fig.~\ref{fig4}(b) for the same and larger values of frustration, 
$J^{\prime\prime}_{Zn}\!=\!2J^{\prime}_{Zn}\!=\!0.1J$
($J^{\prime}_{tot}\!=\!0.4J$). They yield the lower bounds for the 
slopes $R(0)\!\approx\!0.7$ and $0.9$, respectively \cite{comment}. The 
results suggest that a stronger frustration, close to the latter data set 
or somewhat larger, is present in the real Zn-doped CuO$_2$
plane. This is still  
a reasonably modest amount of frustration, well within the window
suggested by the 
three-band model calculations.

Using the same QMC analysis, we have also investigated an alternative
mechanism of the enhanced order suppression by impurities, in which the
strength  of the $J$ bonds in the vicinity of Zn is reduced by the lattice
distortion \cite{Oguchi}. We found that changing the bond strength by
15\% ($\delta \widetilde{J}_{tot}=1.2J$)  
changes the slope R(0) by at most 
a few percent, ruling out the lattice-distortion mechanism of
the order suppression as a viable alternative to our theory.

{\it Conclusions.}---We have proposed that impurity-doped strongly
correlated systems develop significant frustrating interactions,  
which are absent or negligible in the corresponding undoped system,
due to electronic degrees of freedom at the scale less than the Hubbard-$U$. 
Applying this mechanism to the problem of the doped non-frustrated
Heisenberg model,  
relevant to Zn-doped cuprates, we have found this effect to be the key
to resolving  
discrepancies between experiments and earlier theories. Our analytical
and numerical  
results  agree well with each other. We estimate the total amount of
frustration  
to be  $\agt\!0.4J$ per impurity in the Zn-doped CuO$_2$ plane. In
light of our quantitative  
theoretical results, further high-precision experiments at low doping
are called for. 

{\it Outlook.}--Our theory has far-reaching consequences for diluted
AFs and other doped Mott insulators. The character 
of the percolation transition should change as a result of the  
frustrating interactions across the impurities. The impurity-doping of 
spin chains should introduce weaker links between the broken pieces. 
Recent experiments on doped frustrated $J_1$-$J_2$ 
systems \cite{Carretta1} should be affected by the same mechanism. 
The proposed impurity-induced frustrating interactions should persist 
at the finite doping and may produce a pair-breaking mechanism in the doped 
CuO$_2$ planes.

This work was supported by the DOE, grant DE-FG02-04ER46174, 
and by the Research Corporation (S.~L. and A.~L.~C.), NSC and NCTS of
Taiwan  
(C.-W.~L. and Y.-J.~K.), and by the NSF, Grants No.~DMR-0803510
(A.~W.~S.) and No. PHY05-51164 (KITP).  
C.-W.~L. and Y.-J.~K. acknowledge the NCHC for the support
of  
HPC facilities, and A.~W.~S. would like to thank the NCTS of Taiwan
for travel support. 

\null

\end{document}